# An Efficient Data-aided Synchronization in L-DACS1 for Aeronautical Communications

T. H. Pham
School of Computer Science and Engineering
Nanyang Technological University
Singapore 639798
pham_ht@ntu.edu.sg

A. P. Vinod
School of Computer Science and Engineering
Nanyang Technological University
Singapore 639798
ASVinod@ntu.edu.sg

A. S. Madhukumar
School of Computer Science and Engineering
Nanyang Technological University
Singapore 639798
asmadhukumar@ntu.edu.sg

## ABSTRACT
L-band Digital Aeronautical Communication System type-1 (L-DACS1) is an emerging standard that aims at enhancing air traffic management (ATM) by transitioning the traditional analog aeronautical communication systems to the superior and highly efficient digital domain. L-DACS1 employs modern and efficient orthogonal frequency division multiplexing (OFDM) modulation technique to achieve more efficient and higher data rate in comparison to the existing aeronautical communication systems. However, the performance of OFDM systems is very sensitive to synchronization errors. L-DACS1 transmission is in the L-band aeronautical channels that suffer from large interference and large Doppler shifts, which makes the synchronization for L-DACS more challenging. This paper proposes a novel computationally efficient synchronization method for L-DACS1 systems that offers robust performance. Through simulation, the proposed method is shown to provide accurate symbol timing offset (STO) estimation as well as fractional carrier frequency offset (CFO) estimation in a range of aeronautical channels. In particular, it can yield excellent synchronization performance in the face of a large carrier frequency offset.

## 1. INTRODUCTION
Over the past two decades, the air transport industry has experienced continuous growth and the demand for passenger air traffic is forecast to double the current level by about 2025 [1]. The current air transportation systems will not be able to cope with this growth, e.g., already Very High Frequency communication capacity is expected to saturate in Europe by 2020-2025 [2]. Meeting these growing demands require efficient air-to-ground communication systems providing various data of airplanes in real-time. L-DACS is being proposed as a solution that can coexist with legacy L-band systems and aims to explore digital radio techniques to enable efficient communication for next generation global ATM systems [3]. There are two specifications that are being reviewed for L-DACS, type-1 (LDACS-1) and type-2 (LDACS-2). The type-1 (L-DACS1) specification defined by EUROCONTROL [4] is the most promising and mature candidate for final selection. It is intended to be operated in the frequency range that is mainly utilized by aeronautical navigation aids (e.g. the distance measuring equipment (DME), the military Tactical Air Navigation (TACAN)).

L-DACS1 employs OFDM modulation that has been widely employed in various high bit-rate wireless transmission systems (e.g. WiMAX, WiFi, etc.). OFDM is a modern and effective multi-carrier modulation technique with its advantages of combating impulsive noise, robustness to multipath effects and spectral efficiency. However, OFDM performance is sensitive to receiver synchronization. Carrier frequency offset causes inter-carrier interference (ICI) and errors in timing synchronization can lead to inter-symbol interference (ISI) [5]. Therefore, synchronization is critical to the performance of OFDM systems as well as L-DACS1.

Many techniques have been proposed for effective OFDM synchronization in the literature that include both data-aided schemes [6-8], and blind schemes [9], [10]. The latter schemes use inherent structure of the OFDM symbol (i.e. cyclic prefix). However, these methods usually require a large number of OFDM symbols leading to long delay and large computation cost to achieve satisfactory performance. In contrast, the former schemes employ training symbols (i.e. preamble) for either the auto-correlation of received preamble or the cross-correlation between the copy of transmitted preamble and received symbol at the receiver. Auto-correlation based methods for timing synchronization [8] rely on the repetitions of preamble that offer robustness to large CFO and multi-path channel with low computational cost. These methods are suitable for coarse STO and CFO estimation. In the absence of CFO, cross-correlation based methods achieve an excellent timing synchronization performance [6]. However, the performance of these methods degrades significantly due to the present of large CFO. This limits their use to fine timing estimation in synchronization schemes [7], [11], which use auto-correlation for coarse timing and cross-correlation for fine timing estimation. L-DACS1 transmission is operated in L-band aeronautical channels which suffer from large interference and large Doppler shifts. Synchronization of LDACS1 is hardly addressed in literature. To the best of authors' knowledge, there is only one synchronization method that is presented for an optimized L-DACS1 receiver [12]. This method is based on blind schemes leading to requirements of long synchronization time and large computation cost, and it just tolerates a small CFO which is less than one subcarrier spacing.

In this paper, we propose a novel data-aided synchronization method for L-DACS1 in which the preamble of OFDM frames in L-DACS1 are employed. Both frequency offset estimation and timing synchronization are completed within the duration of this preamble. The proposed method introduces correlation-based timing metrics and an effective scheme that is suitable for computationally efficient hardware implementation. The proposed method is robust against large CFO and achieve accurate STO estimation as well as fractional CFO estimation in a range of aeronautical channels.

## 2. LDACS1 SYNCHRONIZATION

OFDM synchronization consists of STO estimation and CFO estimation. STO estimation is to find the first sample of each OFDM symbol to retrieve a complete OFDM symbol for demodulation. CFO estimation determines the frequency mismatch between transmitted samples at transmitter and received samples at receiver. Synchronization can be performed at receiver based on the special symbols, known as preamble, that are sent at the beginning of each physical frame. L-DACS1 is specified with two preamble symbols. The first symbol has a time domain waveform consisting of four identical parts of length $L$, whereas the second symbol is formed with two identical parts of length $2L$. L-DACS1 is specified with $L = 16 * N_{ov}$ where $N_{ov}$ is the oversampling factor in the receiver, which is typically four [4]. The preamble of L-DACS1 in time domain after adding the cyclic prefix ($T_{cp}$) and applying windowing ($T_w$) is depicted in Figure 1.

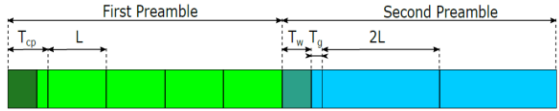

**Figure 1. The structure of the LDACS1 Preamble symbols.**

A signal is transmitted through a frequency-selective channel and corrupted by a zero mean complex white Gaussian noise $\eta_n$. At the receiver, the signal is received and down-converted to baseband for demodulation. The samples of received signal can be expressed as follows:

$$r_n = e^{j(2\pi\epsilon n/N)} \sum_{l=0}^{C-1} h_l x_{n-l} + \eta_n, \quad (1)$$

where $x_n$ and $r_n$ denote transmitted and received samples, respectively. $h_l$ is the baseband equivalent discrete-time channel impulse response of length $C$. $\epsilon$ denotes the normalised carrier frequency offset between transmitter and receiver.

### 2.1 Proposed Timing Metrics

We propose timing metrics that take advantage of L-DACS1 preamble structure and energy correlation. We define two autocorrelation metrics based on the periodic parts of the first preamble symbol as follows:

$$AC1(n) = \sum_{m=0}^{2L-1} c1(m,n),$$
$$AC2(n) = \sum_{m=0}^{2L-1} c2(m,n),$$
$$c1(m,n) = r^*_{n-m} r_{n-m-L},$$
$$c2(m,n) = r^*_{n-m} r_{n-m-2L} \quad (2)$$

where * denotes complex conjugation. The first autocorrelation metric searches for quarter repetition whereas the second finds half repetition of the first preamble symbol. These metrics are employed not only for frame detection but also for accurate CFO estimation which is discussed in following subsection.

An energy metric, $ENE$ is proposed which measures the received symbol energy in a duration of length $2L$. This metric is used as a reference for detect an incoming preamble.

$$ENE(n) = \sum_{m=0}^{2L-1} r^*_{n-m} * r_{n-m} \quad (3)$$

To keep the exposition simple, assume an ideal channel with noise. Then, samples of the received preamble are expressed as follows:

$$r_n = e^{j(2\pi\epsilon n/N)} * p_n + \eta_n, \quad (4)$$

where $p_n$ denotes the transmitted samples of the preamble symbols. The metrics in (2) is derived as follows:

$$AC1(n) = e^{-j\phi_1} \sum_{m=0}^{2L-1} p^2_{n-m} + \eta_1(n), \quad (5)$$

$$\eta_1(n) = \sum_{m=0}^{2L-1} \eta^*_{n-m}\eta_{n-m-L} + p_{n-m}(\eta^*_{n-m} + \eta_{n-m-L})$$

$\eta_1(n)$ presents the noise part on the metric. $\Phi_1 = 2\pi\epsilon(L/N)$ is the phase rotation caused by CFO. Similarly, $AC2(n)$ and $ENE(n)$ can be derived as follows:

$$AC2(n) = e^{-j\phi_2} \sum_{m=0}^{2L-1} p^2_{n-m} + \eta_2(n),$$

$$ENE(n) = \sum_{m=0}^{2L-1} p^2_{n-m} + \eta_e,$$

where $\eta_2(n)$ and $\Phi_2 = 2\pi\epsilon(2L/N)$ are the noise part and phase rotation part of the second autocorrelation metric. $\eta_e$ denotes the noise part of energy metric. If the noise part and phase rotations part are negligible, the autocorrelation metrics equal the energy metric when the first preamble symbol is received.

An energy correlation metric is presented for accurate STO estimation instead of using signal cross-correlation as in conventional approach. The energy correlation is insensitive to phase rotation caused by CFO. Moreover, the metric are computed on real numbers leading to reduce computational requirements in comparison to the signal cross-correlation. The periodic waveform of preamble symbols causes the peaks of cross-correlations. Not only a peak occurs at the correct STO, but also there are secondary peaks which present at the repetitions of waveform. The secondary peaks may cause incorrect STO estimations. The proposed energy correlation metric is expressed as follows:

$$XCR(n) = \sum_{m=0}^{D-1} |c2(m,n)| * a_m \quad (6)$$

The proposed metric uses $|c2(m, n)|$ for energy correlation instead of instant energy $r^*_{n-m} r_{n-m}$ like in [8]. $a_m$ is an energy vector of the transmitted preamble. $D$ is the length of the vector $a_m$ illustrated in Figure 2. It should be noted that when preamble symbols presents at received signal, $|c2(m, n)|$ equals the instant energy. However, $|c2(m, n)|$ eliminates the periodic feature of the preamble as shown in Figure 2 because ISI occurs when received signal multiplies with its delayed version. Therefore, the proposed metric can reduce the secondary peaks leading to improved STO estimation accuracy.

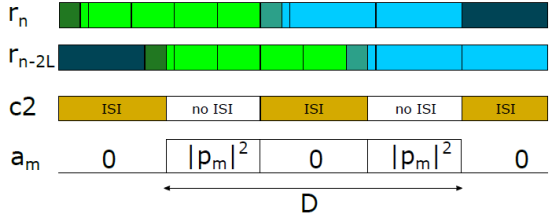

**Figure 2. The proposed metrics on the L-DACS1 preamble.**

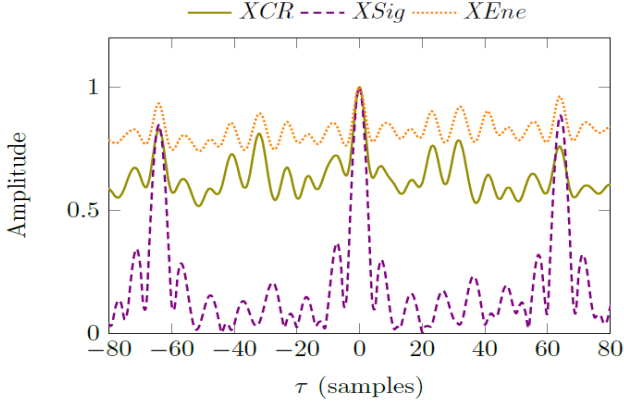

**Figure 3. Cross-correlation metrics for STO estimation at SNR = 10dB.**

Figure 3 shows the normalised values of proposed metric (*XCR*) in comparison to signal cross-correlation (*XSig*) and instant energy correlation (*XEne*). As can be seen, $\tau = 0$ corresponds to the correct STO where the largest peak occurs. The secondary peaks of proposed metric located at -64, 64 (i.e. *-L, L*) is reduced compared to *XSig* and *XEne*.

## 2.2 Proposed Synchronization Flow

The proposed synchronization employs the aforementioned timing metrics to jointly estimate STO and CFO in an efficient synchronization flow as follows: First, incoming preamble symbols are detected by a comparison operator between the autocorrelation metrics with the signal energy metric. Then, STO estimation is performed using the proposed energy correlation metric and CFO is estimated by using the autocorrelation metrics. With the efficient synchronization flow, the proposed method can be easily implemented on hardware as illustrated in Figure 4.

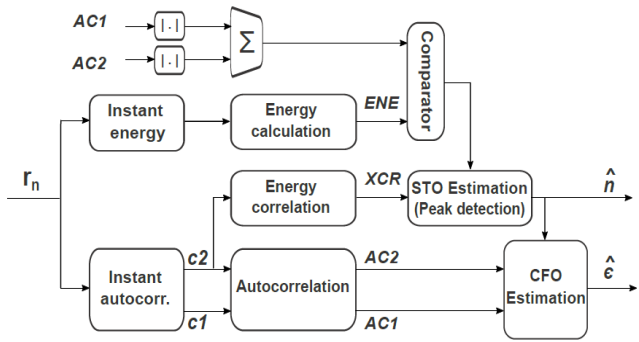

**Figure 4. Proposed scheme for L-DACS1 synchronization implementation.**

The preamble detection is based on the first preamble symbol. When the first symbol presents in received signal, the values of autocorrelation metrics are increased and the condition in (7) will be met.

$$AC(n) > ENE(n), \quad (7)$$

where $AC(n) = |AC1(n)| + |AC2(n)|$. To increase stability, the first preamble symbol is detected when condition (7) is met for m consecutive samples (with $m = 4*N_{oc}$ used throughout this paper). After the first preamble symbol is detected, the synchronization jointly performs STO and CFO estimations.

### 2.2.1 STO estimation

According to (6), when the preamble is received and $c2(m, n)$ matches with $a_m$, XCR will get the largest peak. Therefore, STO can be determined by searching the largest peak of XCR metric. The STO estimation is performed as follows:

$$\hat{n} = \arg\max_{n \in \Delta} XCR(n), \quad (8)$$

where $\hat{n}$ denotes estimated STO and $\Delta$ is a searching window and empirically chosen to equal $56*N_{oc}$.

### 2.2.2 CFO estimation

According to (5), CFO can be estimated as follows:

$$\hat{\epsilon} = \frac{\phi_1}{2\pi}\frac{N}{L} - z_1\frac{N}{L},$$
$$= \frac{\phi_2}{2\pi}\frac{N}{2L} - z_2\frac{N}{2L}, \quad (9)$$

where $\Phi_1, \Phi_2$ are the angle of *AC1, AC2*, respectively. $z_1$ and $z_2$ are an integer. The angle of *AC1* or *AC2* can be used to accurately estimate the CFO if $z_1$ or $z_2$, respectively, equals zero. Because of $-\pi < \Phi_1$; $\Phi_2 < \pi$, the estimation using *AC1* is limited in ± 2 subcarrier spacing, while using *AC2* the estimation range is ± 1 subcarrier spacing, respectively. When CFO is larger than the estimation range, $z_1$ and $z_2$ are non-zero that represent integer CFO that can be estimated as in [13]. Integer CFO estimation is beyond the scope of this paper. CFO estimation using *AC1* has larger range than that using *AC2*. However, the estimation using *AC2* is more accurate compared to *AC1*. Moreover, by using *AC2*, CFO estimation can be performed on both preamble symbol leading to improve estimation accuracy. The proposed method combines both metrics to achieve the CFO estimation with wide range and high accuracy. The proposed CFO estimation is expressed as follows:

$$\hat{\epsilon} = \begin{cases} \frac{\phi_2}{\pi} & \text{if } -\pi/2 < \phi_1 < \pi/2, \\ \frac{\phi_2}{\pi} + 2 & \text{if } \pi/2 < \phi_1, \\ \frac{\phi_2}{\pi} - 2 & \text{otherwise} \end{cases} \quad (10)$$

## 3. SIMULATION

Monte Carlo simulations are performed to evaluate the proposed method for L-DACS1 systems. We investigate the synchronization performance in MATLAB under both AWGN channels and aeronautical propagation channels. 10,000 trials are simulated for each case. The accuracy of synchronization method is measured in terms of fail rate and mean square error (MSE) for STO and CFO estimations, respectively. Fail rate equals the fraction of unsuccessful timing estimations over the total trials (i.e. 10,000 times). MSE is calculated based on the error between CFO estimation and actual CFO.

## 3.1 Performance in AWGN channels

The performance of the proposed method is investigated in comparison to the state of the art (*SoA*) method presented in [13] in AWGN channel with the present of CFO. The comparison is presented in terms of the accuracy of both time synchronization and fractional CFO estimation. The performance of STO estimation is measured in terms of failure rate (%), and the accuracy of CFO estimation is evaluated in terms of mean square error (MSE).

Figure 5 shows the performance of timing synchronization in AWGN channels. *Prop* and *Prop_1.5* denote the proposed methods in cases of CFO absence and large CFO, respectively. The large CFO is set to equal 1.5 subcarrier spacing. The performance of the conventional method in [12] is denoted by *SoA* and *SoA'*. *SoA* is evaluated with the accuracy of coarse STO estimation. This means that a timing synchronization is successful if STO estimation is in cyclic prefix duration. But L-DACS1 requires the accuracy of STO estimation less than 1/11 cyclic prefix length [4]. *SoA'*, *Prop* and *Prop_1.5* is measured for fine STO estimation in which a successful estimation requires an accuracy less than 1/11 cyclic prefix length.

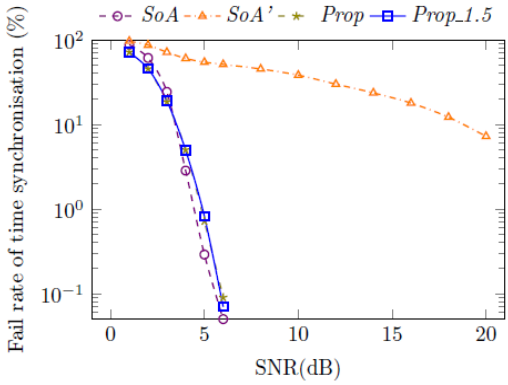

**Figure 5. Performance of time synchronization in AWGN channels with a frequency offset.**

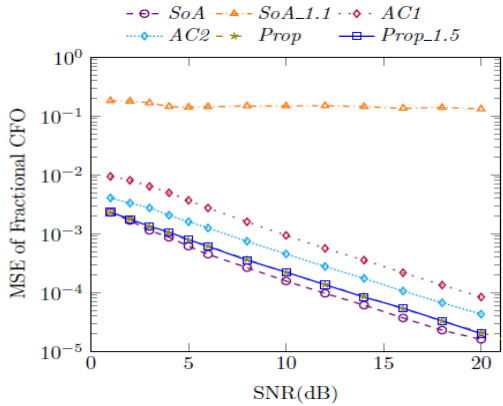

**Figure 6. Performance of frequency offset estimation in AWGN channels.**

As can be seen in Figure 5, the proposed method is robust against a large CFO. *Prop* and *Prop_1.5* have almost identical fail rate. The proposed method achieves an excellent STO estimation with present of large CFO at SNRs above 5 dB. The proposed method outperforms the *SoA'* with the requirement of STO estimation accuracy less than 1/11 cyclic prefix length. The proposed method obtains similar fail rate compared to the SoA. However, the *SoA* accuracy is for coarse STO. A fine STO estimation is required following the *SoA* operation to obtain fine STO. This scheme requires further computation and long estimation time which is up to 0.25 s at high SNRs [12]. The proposed method can obtain the fine STO within the duration of preamble (i.e. 240 us).

As can be seen in Figure 6, the CFO estimation using *AC2* metric is better in comparison to *AC1* metric. The proposed method *Prop* and *SoA* are performed on two preamble symbols and have almost identical performance that is more accurate compared to using *AC2* metric. Moreover, when CFO is large than one subcarrier spacing, the *SoA_1.1* performance has significant degradation, while the proposed method *Prop_1.5* still maintains good performance with CFO = 1.5 subcarrier spacing.

## 3.2 Performance in aeronautical channels

In this subsection, the proposed synchronization method is further investigated in aeronautical channels. The channels are modelled in practical scenarios such as terminal maneuvering (TMA) area and en-route (ENR) flight without/with DME interference. The simulation parameters for channel model are referenced from [12], [14]. The channel models take into account many wireless channel effects including delay spread, Doppler spread, phase noise, and channel interference. The ENR channel was modelled with a strong line-of-sight path and reflected paths with a delay of $\lambda 1 = 0.3$ μs and $\lambda 2 = 15$ μs. The maximal velocity of airplane is assumed 1360 km/h, corresponding to a maximal Doppler shift of 1250 Hz. The TMA channel was modelled with maximum path delay of 10 us, Rician factor of 10 dB and maximal Doppler shift of 624 Hz. A realistic model presented in [15] is used to simulate DME interference for the area around Paris as this is the area with the highest density of DME ground stations in Europe. For this DME interference model, there are three DME interference sources with pulse rate of 3600 pulse pairs per second for each. The first DME channel locates at -0.5 MHz offset to L-DACS1 centre frequency with interference power of -67.9 dBm at L-DACS1 Rx input. The other two channels are at +0.5 MHz offset with interference power of -74 dBm and -90.3 dBm.

Figure 7 and Figure 8 respectively depict the timing synchronization and CFO estimation performance of the proposed method in the mentioned scenarios of aeronautical channels in comparison to its performance in AWGN channels. ENR denotes the synchronization performance in an ENR channel without DME interference while ENR-DME implies the case in an ENR channel with DME interference. TMA denotes the synchronization performance in TMA scenario.

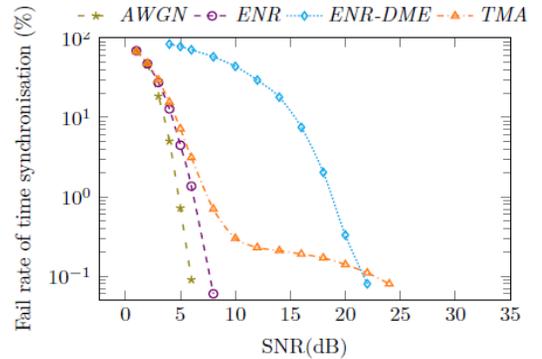

**Figure 7. Time synchronization performance in aeronautical channels.**

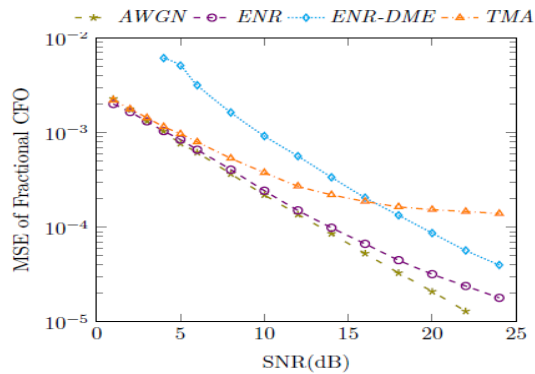

**Figure 8. Frequency offset estimation performance in aeronautical channels.**

In Figure 7 and Figure 8, in case of an ENR channel without DME interference, the accuracy of STO and CFO estimation has a slight decrease compared to the case in an AWGN channel. The proposed method can achieve excellent accuracy in the ENR channel at SNRs above 8 dB. However, when DME interference presents in an ENR channel, the performance of proposed method has considerable degradation. The accuracy of STO estimation and CFO estimation reduce about 10 dB and 4 dB, respectively. The timing synchronization can only achieve good performance at SNRs above 20 dB. In case of the TMA channel, the synchronization has almost similar performance compared to the case of AWGN and ENR channels at SNRs below 8 dB. The CFO accuracy of TMA saturates at SNRs above 10 dB while STO estimation of TMA slightly improve at SNRs above 10 dB and achieve an excellent accuracy at SNRs above 22 dB.

## 4. CONCLUSION

L-DACS1 is being proposed as a solution that can co-exist with legacy L-band systems and aims to explore OFDM-based radio techniques to enable high data rate communication for next-generation global ATM systems. This paper has presented and evaluated a novel synchronization method for L-DAC1. The proposed method is designed to achieve synchronization accuracy and to be much robust to large CFO than the state of the art method. Moreover, the proposed method can obtain good synchronization in a short time within the duration of preamble. The accuracy of the proposed method is also demonstrated under several aeronautical channels. For the future, the synchronization can be further studied with interference mitigation techniques to reduce the effect of DME interference. It is intended to implement the proposed method on a hardware platform such as FPGA. Performing synchronization method on efficient hardware could lead to reducing the power consumption for LDACS1 systems.